\documentclass[pra,twocolumn,superscriptaddress,showpacs]{revtex4}
\usepackage[dvips]{graphicx}
\usepackage{color}
\usepackage{amsmath}
\usepackage{time}

\newcommand{\h}{{\mathbf H}}

\newcommand{\Ad}[1]{{\mathbf a}^\dag_{#1} }
\newcommand{\A}[1]{{\mathbf a}_{#1} }

\begin{document}
%
%
%
\preprint{LA-UR-06-2838}
\title[K-41]
   {A quantitative study of spin noise spectroscopy in a classical gas of $^{41}$K atoms}

\author{Bogdan~Mihaila}
\affiliation{Theoretical Division,
   Los Alamos National Laboratory,
   Los Alamos NM 87545}

\author{Scott~A.~Crooker}
\affiliation{National High Magnetic Field Laboratory,
   Los Alamos National Laboratory,
   Los Alamos NM 87545}

\author{Dwight~G.~Rickel}
\affiliation{National High Magnetic Field Laboratory,
   Los Alamos National Laboratory,
   Los Alamos NM 87545}

\author{Krastan~B.~Blagoev}
\affiliation{Theoretical Division,
   Los Alamos National Laboratory,
   Los Alamos NM 87545}

\author{Peter~B.~Littlewood}
\affiliation{Cavendish Laboratory,
   Madingley Road,
   Cambridge CB3 0HE,
   United Kingdom}

\author{Darryl~L.~Smith}
\affiliation{Theoretical Division,
   Los Alamos National Laboratory,
   Los Alamos NM 87545}

\date{\today, \now}

\begin{abstract}
We present a general derivation of the electron spin noise power
spectrum in alkali gases as measured by optical Faraday rotation,
which applies to both classical gases at high temperatures as well
as ultracold quantum gases. We show that the spin-noise power
spectrum is determined by an electron spin-spin correlation
function, and we find that measurements of the spin-noise power
spectra for a classical gas of $^{41}$K atoms are in good agreement
with the predicted values. Experimental and theoretical spin noise
spectra are directly and quantitatively compared in both
longitudinal and transverse magnetic fields up to the high magnetic
field regime (where Zeeman energies exceed the intrinsic hyperfine
energy splitting of the $^{41}$K ground state).
\end{abstract}


\maketitle


\section{Introduction}

In general, the magnitude of the measured response of a system to an
external perturbation decreases as the size of the system decreases,
and therefore conventional probes based on the measurement of linear
response often become impractical for the study of nanosystems.
However, the fluctuation-dissipation theorem~\cite{Kubo} guarantees
that the linear response of a system can also be determined from the
spectrum of the system's internal fluctuations while in equilibrium.
To probe the intrinsic fluctuations of a system one can use any of a
variety of so-called `noise spectroscopies', which typically disturb
the system much less as compared with measurements based on
intentional perturbation of the system~\cite{Aleksandrov, Sleator,
Awschalom, Weissman, Sorensen, Kuzmich, Smith, Mitsui, Ito,
ref:spin, Oestreich}.  Further, fluctuation signals generally offer
more advantageous scaling with size as the system size is
reduced~\cite{Mamin, ref:spin}.


Recently, we utilized a noise spectroscopy based on ultrasensitive
magneto-optical Faraday rotation applied to classical gases of
alkali atoms~\cite{ref:spin} with an aim to measure the spectrum of
intrinsic spin (magnetization) fluctuations in an atomic ensemble.
In this experiment a linearly polarized laser, detuned from one of
the fundamental \emph{s-p} (D1 or D2) optical transitions of the
alkali atom, was transmitted through the alkali vapor (which itself
was in thermal equilibrium). To leading order, the laser detuning
was sufficiently large to ensure no absorption of the laser by the
atoms, and so the laser was primarily sensitive to the dispersive
(real) part of the atomic dielectric function through the vapor's
spin-sensitive index of refraction~\cite{Happer}. Stochastic spin
fluctuations in the alkali ensemble therefore imparted small
fluctuations in the polarization rotation (Faraday rotation) angle
of the transmitted laser.  These Faraday rotation angle
fluctuations, measured as a function of time, exhibited power
spectra showing clear noise resonances at frequencies corresponding
to the differences between the various hyperfine/Zeeman atomic
levels of the alkali atom.  The experiments in Ref.~\cite{ref:spin}
were performed at relatively high temperatures, where the alkali
atoms behave as a classical Boltzmann gas and interatomic
interactions are unimportant. In addition, these experiments were
conducted in the low magnetic field regime where Zeeman energies
were much less than the typical hyperfine splittings of the atomic
ground state.  From the spin noise spectra alone and with the atomic
ensemble remaining in thermal equilibrium, a determination of the
atomic g-factors, nuclear spin, isotope abundance ratios, hyperfine
splittings, nuclear moments and spin coherence lifetimes was
possible.  These properties of classical alkali atoms are, of
course, already well known~\cite{Corney}; the experiments described
in Ref.~\cite{ref:spin} established the practical value of spin
noise spectroscopy for determining the properties of atomic gases
and the magnitude of spin noise signals expected under various
experimental conditions.

We have also recently suggested~\cite{us} that this type of spin
noise spectroscopy is a promising experimental probe for ultracold
alkali atomic gases because: i)~it is only weakly perturbing and,
ii)~based on the classical alkali gas measurements, large noise
signals are expected. Ultracold gases of alkali
atoms~\cite{anderson,davis,bradley,demarco} provide experimentally
accessible model systems for probing quantum states that manifest
themselves at the macroscopic scale.  Because the temperature is
very low, interactions between the alkali atoms are important, and
novel many-body quantum states arise because of these interactions.
The ability to vary the effective interatomic interaction (by
varying an external magnetic field and thus adjusting the relative
strength of the hyperfine and Zeeman interactions) makes ultracold
atom gases especially interesting model systems for a wide range of
quantum many-body systems.  The properties of ultracold gases of
alkali atoms are of great interest and are just beginning to be
understood.  New experimental probes of these systems aimed at
revealing the underlying interatomic interactions will be very
useful.

Faraday rotation in alkali gases is sensitive to the projection of
the atom's electron spin in the direction of laser propagation
\cite{Aleksandrov, Kuzmich, Happer}. In general, projections of
electron spin alone are not good quantum numbers of the alkali atom
Hamiltonian.  At low magnetic fields where the Zeeman energies are
smaller than or comparable to the hyperfine interaction, the
electron and nuclear spins are entangled and no projection of
electron spin is a good quantum number.  At strong magnetic fields
where the Zeeman splitting is much larger than the hyperfine
splitting, the electron spin projection in the direction of the
magnetic field becomes a good quantum number, but electron spin
projection orthogonal to the magnetic field is not.
Thus, noise spectroscopy at low temperature can be
performed with the magnetic field either parallel or orthogonal to
the direction of laser propagation whereas in the high-field limit,
the magnetic field should be orthogonal to the direction of laser
propagation.

In this paper, we derive a general expression for the spin noise
power spectrum in alkali gases as measured by optical Faraday
rotation and show that the noise power spectrum is determined by an
electron spin-spin correlation function.  This general expression
applies to both classical gases at high temperature as well as
ultracold quantum gases.  We make detailed calculations of the
expected noise spectra for a classical gas of $^{41}$K atoms and
compare these calculated results with quantitative measurements of
the spin noise power.  We find good agreement between the calculated
and measured results in both the low- and high-magnetic field limit.
Isotopically enriched alkali vapors of $^{41}$K were chosen because
the relatively small hyperfine splitting of this atom ($\sim$254
MHz) allows to approach the high-field limit rather easily (magnetic
fields of order 100 Gauss).

The paper is organized as follows: In Sec.~\ref{response} we
derive the general expression for the spin noise power spectrum as
measured by Faraday rotation. In Sec.~\ref{K-41} we consider the
specific case of a classical gas of $^{41}$K atoms and calculate
the expected noise spectra. In Sec.~\ref{experiment} we present
experimental results of spin-noise measurements on a classical gas
of $^{41}$K atoms, and compare with our theoretical results. We
present our conclusions in Sec.~\ref{conclusions}.


\section{Response function for spin noise spectroscopy}
\label{response}

\begin{figure}[b!]
   \includegraphics[width=0.9\columnwidth]{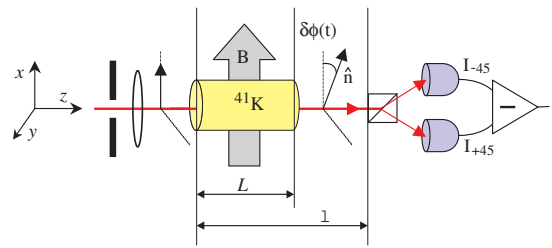}
   \caption{\label{fig:setup}(Color online)
   Experimental schematic showing how electron spin noise in a classical
   vapor of $^{41}$K atoms is probed via optical Faraday rotation.
   Spin fluctuations in the vapor impart Faraday rotation fluctuations $\delta\phi(t)$
   on a linearly-polarized and detuned laser beam, which are measured
   in a sensitive optical bridge. The external magnetic field can be applied
   orthogonal to (as shown) or parallel to the direction of laser propagation.}
\end{figure}

The experimental arrangement for Faraday rotation measurements of
spin noise is shown schematically in Fig.~\ref{fig:setup}.  A
linearly polarized laser, detuned from a \emph{s-p} (D1 or D2)
atomic transition of the alkali atoms, traverses a cell containing
the atomic gas.  A magnetic field is applied either perpendicular
(as shown in Fig.~\ref{fig:setup}) or parallel to the laser
propagation direction. After traversing the atomic gas the laser
beam is divided by a beam splitter at 45$^\circ$ to the original
polarization direction. The split optical beams are detected by a
pair of matched photodiodes and the difference signal (proportional
to Faraday rotation, which is proportional to the magnetization of
the alkali ensemble) is analyzed by a spectrum analyzer.  Full
experimental details are given in Section~\ref{experiment}.

In the electronic ground state of alkali atoms (s-orbital) there is
a strong hyperfine coupling between the nuclear and electron spins.
For the electronic p-orbital the hyperfine splitting is weak because
the p-orbital has a node at the nuclear position, however there is a
strong spin-orbit coupling between the p-orbital and its spin.
Photons directly couple to the spatial part of the electron wave
function, but because of the spin-orbit splitting in the final state
of the optical transition there is an indirect coupling between the
photons and the electron spin. A fluctuating birefringence, that is
a difference in refractive index for left and right hand circular
polarizations, results from fluctuations in the electron spin and
leads to a fluctuating Faraday rotation $\delta\phi(t)$ of the
transmitted laser. The experiment is therefore sensitive to
fluctuations of electron spin projection in the direction of laser
propagation.

The polarization rotation angle noise is
\begin{align}
   \frac{\phi_N(\omega)}{\sqrt{\delta f}} = \ &
   \Bigl [ P(\omega) \Bigr ]^{1/2}
   \>,
\end{align}
where the noise power is
\begin{align}
   P(\omega) = \ &
\label{power_noise}
   \int \mathrm{d}t \ e^{\mathrm{i} \omega t} \
   \langle \delta\phi(t) \, \delta\phi(0) \rangle
   \>.
\end{align}
The optical field is characterized by the vector potential at
position $\mathbf{r}$, which we can write as
\begin{align}
   \mathbf{A}(\mathbf{r},t) = \ &
   \mathbf{a}_0(\mathbf{r},t) \ + \
   \sum_k \ \mathbf{\delta a}_k(\mathbf{r},t)
   \>,
\label{eq:Avec}
\end{align}
where
\begin{align}
   \mathbf{a}_0(\mathbf{r},t) = \ &
   \mathbf{A}_0 \ F(\mathbf{r}) \
   e^{\mathrm{i} \frac{\Omega}{c} (z - c t)}
   \>,
\end{align}
is the incident laser beam and
\begin{align}
   \mathbf{\delta a}_k(\mathbf{r},t) = \ &
   \mathbf{A}_0 \cdot \tilde a_k(\Omega,\theta) \ F(\mathbf{r}_k) \
   e^{\mathrm{i} \frac{\Omega}{c} z_k} \
   \frac{e^{\mathrm{i} \frac{\Omega}{c} \bigl ( | \mathbf{r} - \mathbf{r}_k | - c t \bigr )}}
        {| \mathbf{r} - \mathbf{r}_k |}
   \>
\end{align}
is the scattered optical beam.  Here, $F(\mathbf{r})$ is the beam
profile, normalized to unity in the cross-section plane orthogonal
to the direction of the incident beam, $\hat z$, i.e.
\begin{align}
   1 = & \int \mathrm{d}^2 \rho \ | F(\mathbf{r}) |^2
   \>,
   \qquad \mathrm{with} \ \mathbf{r}=(\vec \rho,z)
   \>.
\end{align}
We denote by $\theta$, the scattering angle between $\hat z$ and
$(\mathbf{r} - \mathbf{r}_k)$, the position of the atom $k$ and
the point $\mathbf{r}$. We introduced $\mathbf{A_0} = A_0\, \hat
\epsilon$, where $A_0$ is a strength parameter and $\hat \epsilon
= \hat x$ is the linear polarization of the incident beam. The
scattering amplitude matrix is
\begin{align}
   \tilde a^{\mathrm{in},\, \mathrm{out}}(\Omega,\theta) = \ &
   \frac{r_0}{m_0} \,
   \frac{1}{\hbar} \,
   \frac{(\mathbf{P} \cdot \hat \epsilon_\mathrm{in}^*) \, \mathcal{P}_J \,
         (\mathbf{P} \cdot \hat \epsilon_\mathrm{out}^*) }
        {\Omega_{J\, R} - \Omega}
   \>,
\end{align}
where $\Omega$ is the angular frequency of the laser,
$r_0=e^2/(mc^2)$ is the classical electron radius, $m_0$ is the
electron mass, $\mathbf{P}$ is the electron momentum operator,
$\mathcal{P}_J =\Sigma |J\rangle \langle J|$ is the projection
operator on to the near-resonant set of intermediate state, $\{|
J\rangle\}$, and $\Omega_{J\, R}$ is the atomic resonance angular
frequency. In Eq.~\eqref{eq:Avec} we have assumed that
$|\mathbf{\delta a}_k(\mathbf{r},t)| \ll
|\mathbf{a}_0(\mathbf{r},t)|$, i.e. we work in the weak scattering
limit.

The polarization beamsplitter forms two beams with intensities
\begin{align}
   I_{\pm}(\mathbf{r}) = \ &
   | \mathbf{A}(\mathbf{r},t) \cdot \hat \epsilon_\pm |^2
   \>,
\end{align}
and linear polarizations
\begin{align}
   \hat \epsilon_\pm = \ &
   \frac{1}{\sqrt 2} \ ( \hat x \pm \hat y )
   \>.
\end{align}
The difference of the two beam intensities in the weak scattering
approximation is
\begin{align}
   I_+(\mathbf{r}) \ - \ & I_-(\mathbf{r})
   \\ \notag = \ &
   2 \, \bigl [ \mathbf{a}_0(\mathbf{r},t) \cdot \hat \epsilon_\pm \bigr ]^* \
   \bigl [
   \mathbf{\delta a}(\mathbf{r},t) \cdot ( \hat \epsilon_+ - \, \hat \epsilon_- )
   \bigr ]
\end{align}
with the components of the vector potential
$\mathbf{A}(\mathbf{r},t)$ evaluated in the forward scattering
direction (Eq.~\eqref{eq:Avec}, with $\theta=0$).  Then
\begin{align}
   \mathbf{a}_0(\mathbf{r},t) \cdot \hat \epsilon_\pm = \ &
   \frac{1}{\sqrt 2} \ A_0 \
   F(\mathbf{r}) \
   e^{\mathrm{i} \frac{\Omega}{c} (z - c t)}
   \>,
\end{align}
where $\mathbf{a}_0(\mathbf{r},t) \cdot \hat \epsilon_+ =
\mathbf{a}_0(\mathbf{r},t) \cdot \hat \epsilon_-$, and
\begin{align}
   \mathbf{\delta a}(\mathbf{r},t) \, \cdot \, & ( \hat \epsilon_+ - \, \hat \epsilon_- ) = \
   A_0 \,
   \sum_k \, F(\mathbf{r}_k) \
   e^{\mathrm{i} \frac{\Omega}{c} z_k} \
   \\ & \notag \times \,
   \Bigl [
        \hat \epsilon
        \cdot
        \tilde a_k(\Omega, 0)
        \cdot
        ( \hat \epsilon_+ - \hat \epsilon_- )
   \Bigr ] \
   \frac{e^{\mathrm{i} \frac{\Omega}{c} \bigl ( | \mathbf{r} - \mathbf{r}_k | - c t \bigr )}}
        {| \mathbf{r} - \mathbf{r}_k |}
   \>.
\end{align}

The photodiode bridge measures the difference of the two
intensities integrated over the laser spot at the detection plane.
The integrated intensity difference measured in a cross-section
plane, $\mathcal{I}$, is
\begin{align} &
   \mathcal{I} \equiv
   \int \mathrm{d}^2 \rho \ \bigl [ I_+(\mathbf{r},t) - I_-(\mathbf{r},t) \bigr ]
\label{intermed}
   \\ \notag & = \
   - \mathrm{i} \, A_0^2 \
   \sum_k \, F(\mathbf{r}_k) \
   e^{\mathrm{i} \frac{\Omega}{c} ( z_k - z ) } \,
   \Bigl [ \tilde a_k^{LL}(\Omega,0) - \tilde a_k^{RR}(\Omega,0) \Bigr ]
   \\ \notag & \qquad \times \,
   \int \mathrm{d}^2 \rho \ F(\vec \rho, z)^* \
   \frac{e^{\mathrm{i} \frac{\Omega}{c} | \mathbf{r} - \mathbf{r}_k |}}
        {| \mathbf{r} - \mathbf{r}_k |}
   \>
\end{align}
with $\mathbf{r}$ evaluated on the detection plane.  Here, $\tilde
a^{LL}$ and the $\tilde a^{RR}$ denote the diagonal elements of
the scattering amplitude matrix $\tilde a$ in the circular
polarization basis,
\begin{align}
   \hat \epsilon_{L,R}
   = \ &
   \frac{1}{\sqrt 2} \ ( \hat x \pm \mathrm{i} \, \hat y )
   \>.
\end{align}
In the circular polarization basis, we have:
\begin{align}
   \hat \epsilon_+ - \hat \epsilon_- = & \
   - \mathrm i \ (\hat \epsilon_L - \hat \epsilon_R)
   \>,
   \\
   \hat \epsilon = & \
   \frac{1}{\sqrt 2} \
   (\hat \epsilon_L + \hat \epsilon_R)
   \>.
\end{align}
At the detection plane, at $z=\ell$, the $\vec \rho$ integral in
Eq.~\eqref{intermed} gives
\begin{align}
   \int \mathrm{d}^2 \rho \ F(\vec \rho, \ell)^* & \
   \frac{e^{\mathrm{i} \frac{\Omega}{c} \sqrt{ |\vec \rho - \vec \rho_k|^2 + ( \ell - z_k )^2 }}}
        {\sqrt{ |\vec \rho - \vec \rho_k|^2 + ( \ell - z_k )^2 }}
   \\ \notag &
   \approx \
   2\pi \, \mathrm{i} \ \frac{c}{\Omega} \
   e^{\mathrm{i} \frac{\Omega}{c} ( \ell - z_k ) } \
   F(\mathbf{r}_k)^*
   \>,
\end{align}
where we have assumed that $(\ell - z_k)^2 \gg (\rho - \rho_k)^2$,
and $F(\mathbf{r}_k)$ is slowly varying on the scale of the
optical wavelength.  Thus, we obtain
\begin{align}
   \mathcal{I} = \
   \frac{2\pi c}{\Omega} \
   A_0^2 \
   \sum_k \ |F(\mathbf{r}_k)|^2 \,
            \Bigl [ \tilde a_k^{LL}(\Omega,0) - \tilde a_k^{RR}(\Omega,0) \Bigr ]
   \>,
\end{align}
with
\begin{align}
   \tilde a^{LL}(\Omega,0) - \, \tilde a^{RR}(\Omega,0) = \ &
   \pm \frac{2 r_0}{3 m_0} \, \frac{1}{\hbar} \
   \frac{ |\langle S|p_x|P_x \rangle| ^2}{|\Omega_{J\, R}-\Omega|} \
   \sigma_z
   \>.
\end{align}
Here, $<S|p_x|P_x>$ is the momentum matrix element for the optical
transition, $\hat z$ is the direction of laser propagation, and
the $\pm$ correspond to the resonances $|J\rangle=|L+\frac{1}{2};
I\rangle$ and $|J\rangle=|L-\frac{1}{2}; I\rangle$ states with
$L=1$, respectively. We obtain
\begin{align}
   \mathcal{I} = \ \pm \, &
   \frac{4\pi}{3} \, \frac{c r_0}{m_0} \, \frac{1}{\hbar \Omega} \
   A_0^2 \
   \frac{| \langle S | p_x | P_x \rangle |^2}{\Omega_{J\, R} - \Omega}
   \notag \\ & \times
   \int \mathrm{d}^3 R \
        |F(\mathbf{R})|^2 \ \sigma_z(\mathbf{R})
   \>.
\end{align}
where $\sigma_z(\mathbf{R})$ denotes the electron spin density
operator
\begin{align}
   \sigma_z(\mathbf{R}) = \ &
   \sum_k \ \sigma_{z}^{k} \ \delta(\mathbf{R} - \mathbf{r}_k)
   \>.
\end{align}

The integrated intensity in the detection plane is proportional to
the rotation of the polarization angle. We have
\begin{align}
   \delta \phi = \frac{\mathcal{I}}{2 \, A_0^2}
   \>,
\end{align}
or
\begin{align}
   \delta \phi = \ \pm \, &
   \frac{2\pi}{3} \, \frac{c r_0}{m_0} \, \frac{1}{\hbar \Omega} \
   \frac{| \langle S | p_x | P_x \rangle |^2}{\Omega_{J\, R} - \Omega}
   \\ \notag & \times
   \int \mathrm{d}^3 R \
        |F(\mathbf{R})|^2 \ \sigma_z(\mathbf{R})
   \>.
\end{align}
Then, the noise power is calculated as the two point-correlation
function introduced in Eq.~\eqref{power_noise}
\begin{align}
   P(\omega) = \ &
   \Bigl [
   \frac{2\pi}{3} \, \frac{c r_0}{m_0} \, \frac{1}{\hbar \Omega} \
   \frac{| \langle S | p_x | P_x \rangle |^2}
        {\Omega_{J\, R} - \Omega}
   \Bigr ]^2
   \\ \notag & \times
   \int \mathrm{d}^3 R_1 \
        |F(\mathbf{R}_1)|^2
   \int \mathrm{d}^3 R_2 \
        |F(\mathbf{R}_2)|^2
   \\ \notag & \times
   \int \mathrm{d}t \ e^{\mathrm{i} \omega t} \
   \langle \sigma_z(\mathbf{R}_1,t) \, \sigma_z(\mathbf{R}_2,0) \rangle
   \>.
\end{align}
For a slowly varying beam profile, $| F(\mathbf{R}) |^2$, and a
spatially uniform atomic gas the above equation becomes
\begin{align}
   P& (\omega) = \
   \Bigl [
   \frac{2\pi}{3} \, \frac{c r_0}{m_0} \, \frac{1}{\hbar \Omega} \
   \frac{| \langle S | p_x | P_x \rangle |^2}
        {\Omega_{J\, R} - \Omega}
   \Bigr ]^2
   \\ \notag & \times
   \int \mathrm{d}^3 R \,
        |F(\mathbf{R})|^4
   \int \mathrm{d}t \, e^{\mathrm{i} \omega t}
   \int \mathrm{d}^3 r \,
   \langle \sigma_z(\mathbf{r},t) \, \sigma_z(0,0) \rangle
   \>,
\end{align}
where $\mathbf{R}$ and $\mathbf{r}$ are the center and relative
coordinate combinations of $\{\mathbf{R}_1, \mathbf{R}_2\}$. For a
slowly varying beam profile, $|F(\mathbf{R})|^2$, the $\mathbf{R}$
and $\mathbf{r}$ integrals factorize and
\begin{align}
   \int \mathrm{d}^3 R \
        |F(\mathbf{R})|^4
   = \ & \frac{L}{A}
   \>,
\end{align}
where $L$ is the length of the gas cell, and $A$ denotes the
optical beam area. For a Gaussian beam profile, i.e.
\begin{equation}
   | F(\vec \rho,z) |^2
   \ = \
   \frac{1}{\pi \, {R_0}^2} \, \exp \bigl ( - \rho^2/R_0^2 \, \bigr )
   \>,
\end{equation}
we have
\begin{align}
   A = \ &
   2\pi \, {R_0}^2
   \>
\end{align}
where $R_0$ is the radius at which the beam intensity drops to
$1/e$ of its peak value.

In summary, the polarization rotation angle noise is
\begin{equation}
   \frac{\phi_N(\omega)}{\sqrt{\delta f}} \ = \
   C \
     \Bigl [ \,
       \frac {L \, \rho_0}{A} \
       S(\omega) \,
     \Bigr ]^{1/2}
   \>,
\label{power}
\end{equation}
where
\begin{equation}
    C \ = \
        \frac{2\pi}{3} \
        \frac{c \, r_0}{m_0} \
        \frac{1}{\hbar \Omega} \
        \frac{ |\langle S|p_x|P_x \rangle| ^2}{|\Omega_{J\, R}-\Omega|}
   \>,
\end{equation}
and $S(\omega)$ is the electron spin correlation function
\begin{equation}
    S(\omega) \ = \
    \frac{1}{\rho_0} \ \int \mathrm{d}t \ e^{i \omega t}
    \int \mathrm{d}^3 r \
    \langle \sigma_z({\bf{r}},t) \sigma_z(0,0) \rangle
   \>.
\label{s_omega}
\end{equation}
Here, $\rho_0$ is the density of atoms in the system and
$S(\omega)$ satisfies the sum rule
\begin{equation}
   \int \frac{\mathrm{d}\omega}{2\pi} \ S(\omega) \ = \ 1 \>.
   \label{sum_rule}
\end{equation}

Equations (29) and (30) show that the noise signal decreases linearly
with inverse frequency detuning from the optical resonance. By
contrast, the energy dissipated into the atomic system, either by
optical absorption or Raman scattering, decreases quadratically with
inverse frequency detuning.  Thus noise spectroscopy measurements
are only weakly perturbative in the sense that the noise
spectroscopy signal decreases more slowly with inverse frequency
detuning than does the energy dissipated into the system.


\section{Spin noise for a classical gas of $^{41}$K atoms}
\label{K-41}

The spin noise spectrum consists of a series of resonances
occurring at frequencies corresponding to the difference between
hyperfine/Zeeman atomic levels.  The integrated strength of the
lines gives information about the occupation of the atomic levels
and the one-atom electron spin matrix elements, while the line
shapes depend on the properties of the many body atomic state,
\begin{align}
   S(\omega) =
       \sum_{ij} \
       | \langle i | \mathbf{\hat n} \cdot \sigma | j \rangle |^2 \
       S^{i \rightarrow j}(\omega)
\label{P_omega_ij}
   \>,
\end{align}
where $\mathbf{\hat n}$ is the optical polarization vector,
$\{i,j\}$ label the single atom spin states, $| \langle i |
\mathbf{\hat n} \cdot \sigma | j \rangle |^2$ is a one-atom
electron spin matrix element that determines line strengths and
selection rules, and $S^{i \rightarrow j}(\omega)$ contains
information about the many-body atomic state.

\begin{figure}[b]
   \includegraphics[width=0.9\columnwidth]{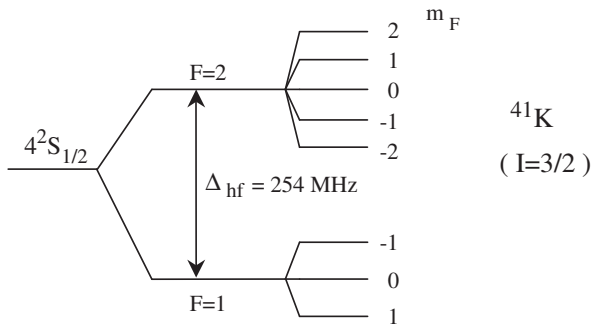}
   \caption{\label{fig:spectrum}
   Schematic of the Zeeman/hyperfine structure of $^{41}$K.
   (For $^{41}$K atoms, $I=3/2$ and $g_I$=0.215.)
   The total angular momentum $F=I+J$ and its projection $M_F$
   are good quantum numbers at $B=0$, and can also be used to unambiguously
   label the atomic levels when $B \neq 0$.
   }
\end{figure}

\begin{figure}[t!]
   \includegraphics[width=0.9\columnwidth]{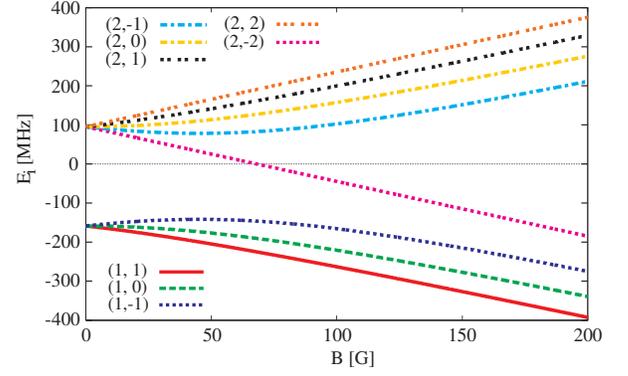}
   \caption{\label{fig:erg}(Color online)
   Energies of the Zeeman-split $^{41}$K atomic levels as a function of magnetic field.
   The levels are labeled by their $(F,M_F)$
   quantum numbers when $B=0$.
   }
\end{figure}


\subsection{Single Atom Electron Spin Matrix Elements}

Alkali atoms are one-electron atoms in the sense that they have
one comparatively weakly bound $s$-electron and a closed-shell
electron core.  Excitations of the closed-shell electron core
occur at a much higher energy scale than the probes that are
considered here.  The atomic levels are eigenstates of the
Hamiltonian of an atom (with nuclear spin $I$) interacting with an
electron (spin $s=\frac{1}{2}$)
\begin{align}
   \h^{(1)} \ = \
  {\frac{\mathbf{p}^2}{2m}} \ + \ \mathbf{V}_{a-e}
   \ = \
   \sum \ H^{(1)}_{k, i j} \ \Ad{k i} \A{k j}
   \>,
\end{align}
where $\mathbf{p}^2/(2m)$ is the kinetic energy, $\mathbf{V}_{a-e}$
is the atom-electron potential
\begin{align}
   \mathbf{V}_{a-e}
   \ = \
   A \ \vec s \cdot \vec i
   \ + \ \vec B \cdot \Bigl ( 2 \mu_e \vec s - g_I \mu_n \vec i \Bigr )
   \>,
\end{align}
with $\mu_e = g_e \mu_B / 2$, with $g_e$ = 2.0023, and $\mu_n =
\mu_B / 1836$. Here, $A$ denotes the strength of the hyperfine
interaction, and the hyperfine splitting is
$\Delta_{\rm{hf}}=(I+\frac{1}{2})A$.

The atom-level wave functions involve both nuclear and electron
degrees of freedom.  In the representation of nuclear and electron
spins the single-particle state are
\begin{equation}
   | i \rangle = 
                 | I_i M_i \rangle \ | s_i m_i \rangle
   \>,
\end{equation}
with $s_i=1/2$. The matrix elements of the one-body Hamiltonian
are
\begin{align}
   & \langle \psi_i | H^{(1)} | \psi_j \rangle
   \ = \
   \delta_{M_i M_j} \delta_{m_i m_j} \
   \\ \notag & \times
   \Bigl [
   \varepsilon(k)
   \ + \
   \frac{A}{2} \ M_j \, (-)^j
   \ + \
   \bigl ( \mu_e \, (-)^j - g_I \mu_n \, M_j \bigr ) \
   B
   \Bigr ]
\label{eq:one-body}
   \\ \notag &
   \ + \
   \frac{A}{2} \
   \delta_{M_i,M_j-1} \ \delta_{m_i \uparrow} \delta_{m_j \downarrow} \
   \Bigl [ I ( I + 1) - M_j ( M_j - 1 ) \Bigr ]^{1/2}
   \\ \notag &
   \ + \
   \frac{A}{2} \
   \delta_{M_i,M_j+1} \ \delta_{m_i \downarrow} \delta_{m_j \uparrow} \
   \Bigl [ I ( I + 1) - M_j ( M_j + 1 ) \Bigr ]^{1/2}
   \>.
\end{align}
Here we denote by $j=1$ the Dirac electron spinor
$|\downarrow\rangle$, and by $j=2$ the spinor $|\uparrow\rangle$.

The one-body Hamiltonian is diagonalized to obtain the atom-level
states, $|\phi_i \rangle$
\begin{equation}
   | \phi_i \rangle
   \ = \ \sum_k \ \alpha_{i k} \ | \psi_k \rangle \>.
\end{equation}
We calculate the electron spin matrix elements
\begin{align}
   \langle \phi_i | & \mathbf{\hat n} \cdot \sigma | \phi_j \rangle
   \ = \
   \sum_{k l} \ \alpha_{i k} \ \alpha_{j l} \
   \\ \notag & \times \
   \Bigl [
   \sin \theta \
   \langle \psi_k | (\sigma_+ + \ \sigma_-) | \psi_l \rangle
   \ + \
   \cos \theta \
   \langle \psi_k | \sigma_0 | \psi_l \rangle
   \Bigr ]
   \>,
\end{align}
where $\mathbf{\hat n} = (\sin \theta, 0, \cos \theta)$.

\begin{figure}
   \includegraphics[width=0.9\columnwidth]{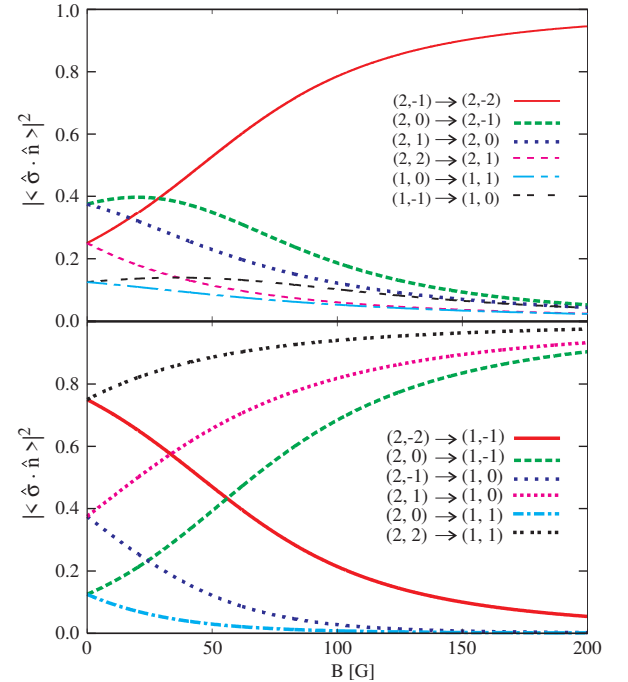}
   \caption{\label{fig:Sampl_90}(Color online)
   Squared electron spin matrix elements for the case of transverse magnetic fields ($\theta=90^\circ$).
   The selection rule in this case is $\Delta M_F=\pm 1$.
   The top panel depicts the case of $\Delta F=0$. Because the nuclear moment of $^{41}$K is very small,
   the following pairs of transitions are nearly degenerate in energy (frequency):
   $(2,0)\rightarrow(2,-1)$ and $(1,-1)\rightarrow(1,0)$;  $(2,1)\rightarrow(2,0)$ and
   $(1,0)\rightarrow(1,1)$.  The bottom panel shows transitions for which $\Delta F=\pm 1$.
   The following pairs of transitions are nearly degenerate in energy (frequency):
   $(2,1)\rightarrow(1,0)$ and $(2,0)\rightarrow(1,1)$; $(2,0)\rightarrow(1,-1)$ and
   $(2,-1)\rightarrow(1,0)$.
   }
\end{figure}

\begin{figure}[t!]
   \includegraphics[width=0.9\columnwidth]{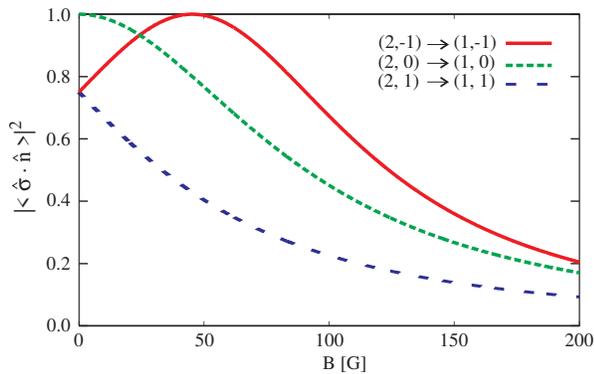}
   \caption{\label{fig:Sampl_0}(Color online)
   Squared electron spin matrix elements in $^{41}$K
   for the case of longitudinal magnetic fields ($\theta=0^\circ$).
   The selection rule in this case is $\Delta M_F=0$.
   }
\end{figure}

\begin{figure}[t]
\includegraphics[width=0.9\columnwidth]{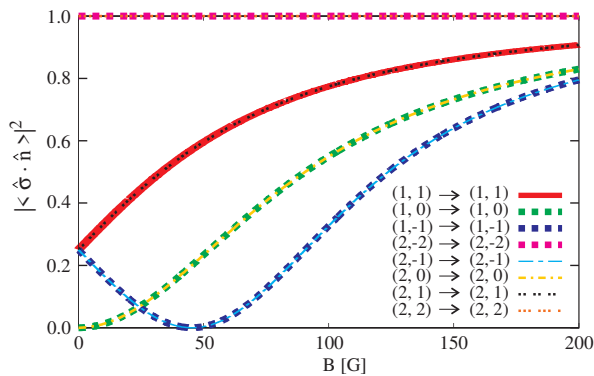}
   \caption{\label{fig:Sampl_0self}(Color online)
   Squared electron spin matrix elements in $^{41}$K for the self-excitation
   transitions (longitudinal magnetic fields, $\theta=0^\circ$).
   These transitions all appear at zero frequency.
   The self-excitation amplitudes of levels $(1,1)$ and $(2,1)$ are equal,
   as are the amplitudes for levels $(1,0)$ and $(2,0)$, and also for $(1,-1)$
   and $(2,-1)$, respectively.
   The self-excitation amplitudes of $(2,-2)$ and $(2,2)$ are both equal to~$1$.
   }
\end{figure}

\begin{figure*}[t]
\includegraphics[width=0.9\textwidth]{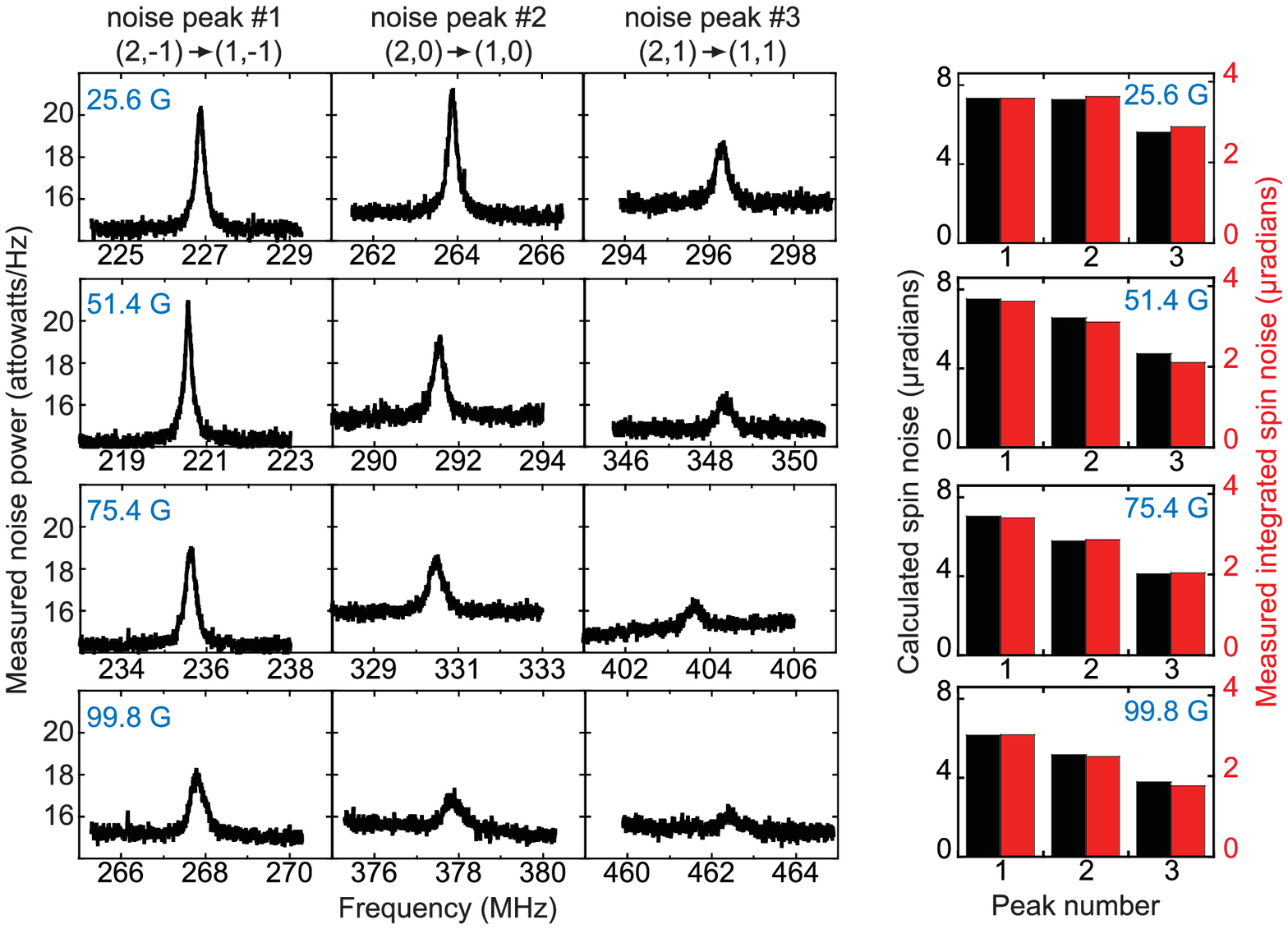}
\caption{\label{fig7}(Color online) The experimentally-measured noise power spectrum
from a gas of $^{41}$K atoms, for longitudinal magnetic fields
$B_L$ ($\theta=0^\circ$). Here, the measured noise \emph{power} density (aW/Hz) is
proportional to the \emph{square} of Faraday rotation noise density (which has units
of radians/Hz$^{1/2}$).  The three allowed spin noise resonances
((2,-1)$\rightarrow$(1,-1), (2,0)$\rightarrow$(1,0), and
(2,1)$\rightarrow$(1,1)) are measured at $B_L$ = 25.6, 51.4, 75.4,
and 99.8 Gauss. Electronic detector and amplifier noise is not
subtracted from this data. At the right, histograms compare the
theoretically calculated spin noise (black bars) with the measured
integrated spin noise (red bars), in units of Faraday rotation (microradians). The
relative measured spin noise agrees well with calculation at all $B_L$. There is an
overall (absolute) discrepancy of approximately a factor of two between theory and
experiment.}
\end{figure*}

For $^{41}$K atoms the nuclear spin is $I=\frac{3}{2}$, and the
hyperfine splitting is $\Delta_{\rm{hf}}=2A$. The schematic of the
Zeeman/hyperfine spectrum for $^{41}$K is shown in
Fig.~\ref{fig:spectrum}. The atom levels are obtained by
diagonalizing the one-body Hamiltonian $H^{(1)}$, and the
atom-level energies are plotted in Fig.~\ref{fig:erg} as a
function of applied magnetic field $B$.  At low magnetic fields
where the Zeeman energies are smaller than or comparable to the
hyperfine interaction, the electron and nuclear spins are
entangled and no projection of electron spin is a good quantum
number.  At strong magnetic fields where the Zeeman splitting is
much larger than the hyperfine splitting, the electron spin
projection in the direction of the magnetic field becomes a good
quantum number.

For transverse magnetic fields ($\theta=90^\circ$), the angular
momentum selection rule is $\Delta M_F=\pm 1$, and the magnetic
field dependence of the transition amplitudes is shown in
Fig.~\ref{fig:Sampl_90}.  At low magnetic fields, there is
strength for all of the $\Delta M_F=\pm 1$ transitions, but at
high magnetic field there is strength only for those transitions
corresponding to a change in the electron spin projection in the
direction of the magnetic field.  For longitudinal magnetic fields
($\theta=0^\circ$), the angular momentum selection rule is $\Delta
M_F=0$, and the $B$ dependence of the corresponding transition
amplitudes is illustrated in Fig.~\ref{fig:Sampl_0}. In
Fig.~\ref{fig:Sampl_0self}, we show the $B$ dependence of the
self-transition amplitudes, present when $\theta=0^\circ$. At low
magnetic fields, there is strength for all of the $\Delta M_F=0$
transitions, but at high magnetic field there is strength only for
the zero frequency self transitions. The zero frequency
self-transitions are hard to measure experimentally due to
environmental noise sources, but they are necessary for completing
the sum rule Eq. (32).


\subsection{Spin correlation function}

In second-quantization notation the electron-spin correlation function is
\begin{align}
   \langle \sigma_z({\bf{r}},t) \sigma_z(0,0) \rangle
   =
   \sum_{ijmn} &
   \langle i | \mathbf{\hat n} \cdot \sigma | j \rangle
   \langle m | \mathbf{\hat n} \cdot \sigma | n \rangle
   \\ \notag & \times \
   \langle
   e^{\mathrm{i}\h t}
   \Ad{i} \A{j}
   e^{- \mathrm{i}\h t}
   \Ad{m} \A{n} \rangle
   \>,
\label{S_2nd}
\end{align}
where $\langle O \rangle$ denotes a thermal average of the operator
$O$. For a non-interacting system of atoms, $\h = \h^{(1)}$, and the
electron-spin correlation function becomes
\begin{align}
   S(\omega)
   =
   \frac{1}{\rho_0}
   \sum_{ij} & \
       | \langle i | \mathbf{\hat n} \cdot \sigma | j \rangle |^2 \
   \delta(\omega - \delta_{ij})
   \\ \notag & \times
   \int \frac{\mathrm{d}^3k}{(2\pi)^3} \
   \rho_{k,i} \, ( 1 - \rho_{k,j} ) \
   \>,
\end{align}
where $\delta_{ij}$ denotes the transition energy between atomic
levels $i$ and $j$ (i.e. $\delta_{ij}=E_{j}-E_{i}$), and we have
introduced the density matrix $\rho_{k,i}= \langle
\Ad{\mathbf{k}i}\A{\mathbf{k}i} \rangle $, describing the
occupation of level $i$, corresponding to momentum $k$. In the
classical (high-temperature) limit, the occupation numbers are
given by the Boltzmann distribution function.  The product of two
occupations, $\rho_{k,i}\, \rho_{k,j}$, can be neglected when
compared to $\rho_{k,i}$ for a Boltzmann gas.  In the high
temperature limit all hyperfine/Zeeman states are equally
populated and  $S^{i \rightarrow j}(\omega)$ is given by
\begin{equation}
   S^{i \rightarrow j}(\omega)
   =
   \frac{1}{N \rho_0} \,
   \delta(\omega - \delta_{ij})
   \>,
\end{equation}
where $N$ is the number of levels in the hyperfine spectrum of the
alkali atom. The $\delta$-functions can be broadened by, for
example, finite spin lifetime effects or the time it takes the
atoms to traverse across the laser beam.

\begin{figure}
\includegraphics[width=0.9\columnwidth]{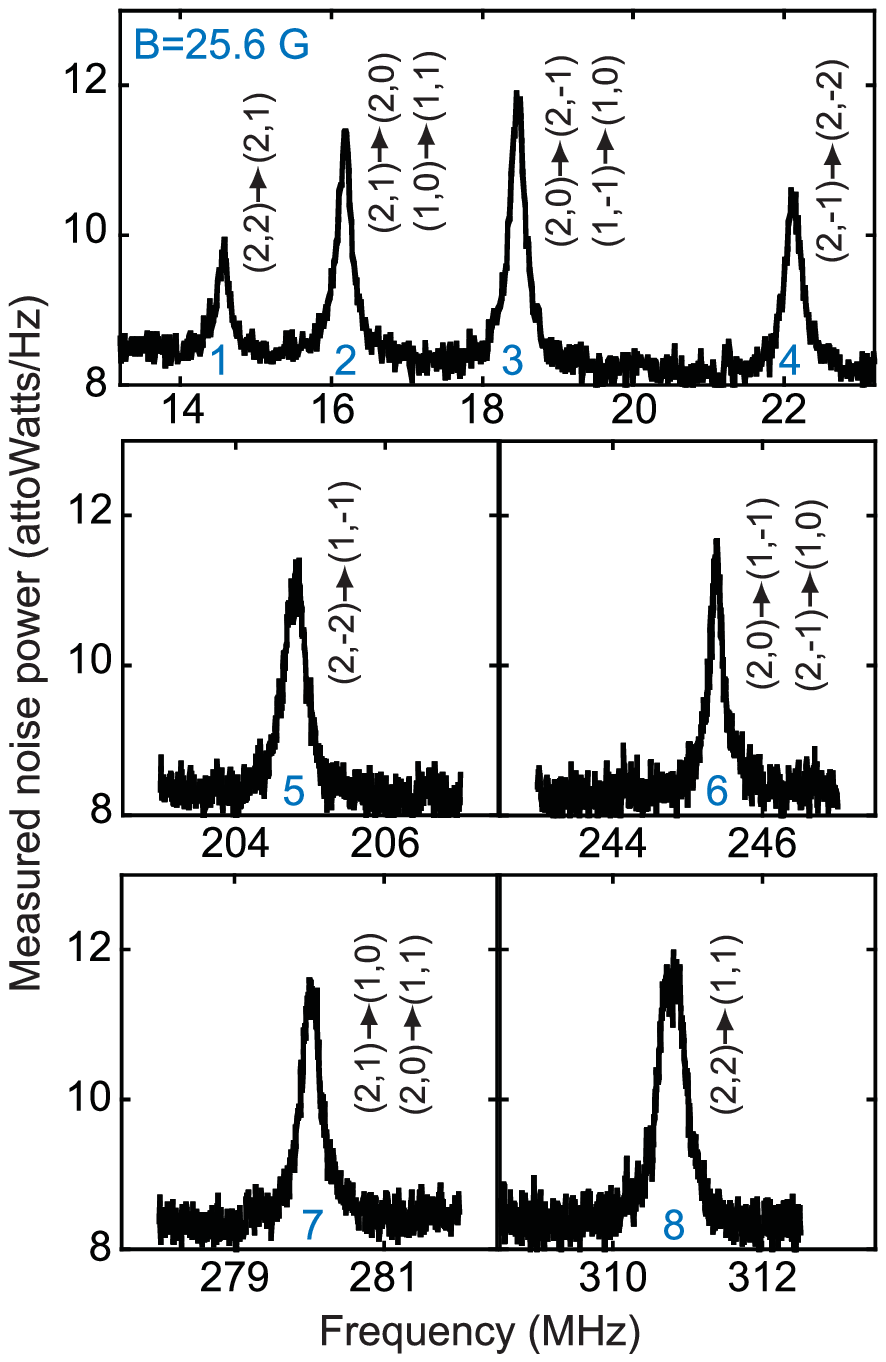}
\caption{\label{fig8}(Color online) The experimentally-measured noise power spectrum
in a gas of $^{41}$K atoms, at a transverse magnetic field of
$B_T$=25.6 G ($\theta=90^\circ$). There are twelve allowed inter-
and intra-hyperfine spin noise resonances.  Four pairs of resonances are nearly
degenerate in frequency, giving eight resolvable spin noise peaks in
the data (peaks are labeled 1-8 as shown).  Background electronic noise from the
detectors and
amplifiers has been subtracted.}
\end{figure}

\begin{figure}[b]
\includegraphics[width=0.9\columnwidth]{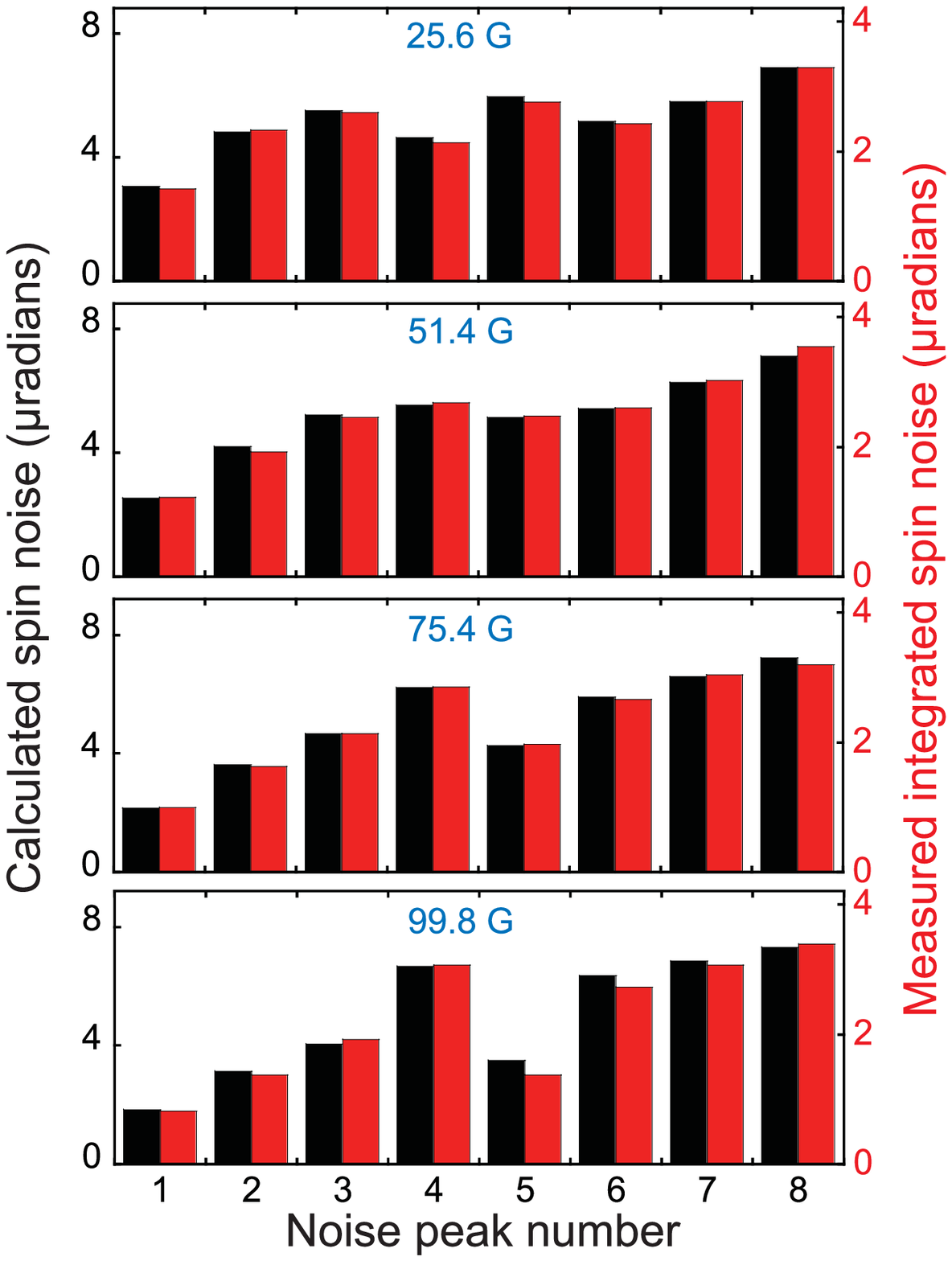}
\caption{\label{fig9}(Color online) Histograms comparing the
calculated spin noise (black bars) with the measured
integrated spin noise (red bars) for all eight resolved noise
resonances in transverse magnetic fields $B_T$ ($\theta=90^\circ$), in units of
Faraday rotation (microradians). Peaks are labeled in Figure 8. The relative
measured spin noise agrees well with
calculation for all values of $B_T$ (25.6, 51.4, 75.4, and 99.8
Gauss). There is an overall (absolute) discrepancy of approximately a factor of
two between theory and experiment. }
\end{figure}


\section{Experimental results}
\label{experiment}

In this section we compare the results of our theoretical model with
experimental data.  We use a gas of the isotope $^{41}$K primarily
because $^{41}$K has a very small hyperfine splitting (254 MHz),
permitting easy access to the ``high magnetic field regime" where
the characteristic Zeeman energies approach and exceed the hyperfine
energy.  Moreover, spontaneous noise resonances in $^{41}$K occur at
relatively low frequencies ($<$500 MHz), where photodetectors are
generally more sensitive. We use a 1~cm long glass vapor cell
containing isotopically-enriched $^{41}$K metal. The cell is
typically heated to 184~$^\circ$C, giving a particle density of 7.3
$\times$10$^{13}$/cm$^3$ in the vapor. The 4 mW probe laser beam,
derived from a continuous-wave Ti:Sapphire laser, is typically
detuned by 100 GHz from the D1 transition (770 nm) of $^{41}$K, and
Faraday rotation fluctuations on the transmitted probe laser beam
are detected by fast balanced photodiodes (New Focus model 1607,
which has 650 MHz bandwidth and 350 V/W gain).  The resulting noise
power spectrum is detected by a 500 MHz spectrum analyzer (Agilent
model 4395). The detectors and amplifiers typically contribute a
frequency-dependent noise density between 4-7 $\times 10^{-18}$
Watts/Hz (4-7 aW/Hz). Using 4 mW of probe laser power, photon shot
noise contributes an additional 8-9 aW/Hz of measured noise. This
measured value of photon shot noise varies somewhat with frequency
because the gain and sensitivity of the detector/spectrum analyzer
combination is not uniform across the entire 0-500 MHz range (in
particular, it falls at the highest frequencies). All values of
measured spin noise from the $^{41}$K atoms include this
frequency-dependent correction.

Figure~\ref{fig7} shows the measured noise power spectrum from
$^{41}$K atoms for the case of longitudinal magnetic fields
($\theta=0^\circ$). These raw data are plotted in units of power
spectral density (aW/Hz) as measured by the spectrum analyzer. In
these experiments, spin fluctuations lead to Faraday rotation
fluctuations, which directly generate voltage fluctuations at the
output of the photodiode bridge.  Figure 7 shows the \emph{power}
spectrum of these voltage fluctuations (proportional to voltage
squared).  As such, these raw data convey the \emph{square} of the
spin noise (or Faraday rotation) spectral density (which itself is
expressible in units of radians/Hz$^{1/2}$). The integrated power within the first noise peak of Figure 7 ((2,-1)$\rightarrow$(1,-1) at 25.6 G) is 2.0 pW, which corresponds to 10 $\mu$V of integrated voltage noise (all instruments have 50 ohm impedance).  Given the 350 V/W detector sensitivity and the total optical power in the probe beam (4 mW), the integrated Faraday rotation noise for this particular spin noise resonance is 3.57 $\mu$rad.

Three spontaneous noise resonances are observable in this
configuration: (2,-1)$\rightarrow$(1,-1), (2,0)$\rightarrow$(1,0),
and (2,1)$\rightarrow$(1,1).  Raw data for all three spin noise
peaks are shown, at four different values of applied field (25.6,
51.4, 75.4, and 99.8 Gauss).  On the right-hand side of the
Figure, the theoretically calculated spin noise of each resonance
(black bars) is compared with the integrated spin noise under each
of the three measured peaks (red bars). Values of spin noise are
expressed in units of Faraday rotation (microradians). In addition
to the electron spin matrix elements, the calculated values take
into account the overall prefactor which depends on laser
detuning, power, beam size, and atom density, per Eq. (29). At any
given field, the \emph{relative} noise contained within the three
peaks agrees very well with calculation.  The absolute calculated
results are approximately a factor of 2 larger than the measured
results (see the scales on the histograms). As the magnetic field
increases beyond 50 G and the calculated spin noise in each of the
three peaks decreases (as anticipated in Fig.~\ref{fig:Sampl_0}),
the measured spin noise (red bars) also decreases, continuing to
show good agreement with theory as the high-field regime is
approached.

For transverse magnetic fields ($\theta=90^\circ$), twelve inter-
and intra-hyperfine spin noise resonances are allowed. For
brevity, Figure~\ref{fig8} shows the actual measured data only for
the case of $B_T$=25.6 G. At $B_T$=25.6 G, eight distinct spin
noise peaks are observed (because the nuclear moment of $^{41}$K
is very small, four pairs of allowed noise resonances are nearly
degenerate in frequency and cannot be individually resolved -- see
the labeling of the peaks in Fig.~\ref{fig8}).  Figure~\ref{fig9}
shows calculated spin noise (black bars) and experimentally
measured integrated spin noise for each of the eight resolved
noise resonances, at $B_T$=25.6, 51.4, 75.4, and 99.8 Gauss (peak
numbers are labeled in Figure 8). Again, the relative spin noise
contained within these peaks agrees very well with theoretical
prediction, but the theoretical results are approximately a factor
of 2 larger than the experimental results in absolute magnitude.
As the magnetic field is increased from the low to the high-field
limit, the relative spin noise that is measured by experiment
changes in accord with the calculations shown in Figure 4.


\section{Summary and conclusions}
\label{conclusions}

In summary, we have derived a general expression for the electron
spin noise power spectrum in alkali gases as measured by Faraday
rotation. We have shown that the noise power spectrum is
determined by an electron spin-spin correlation function. A
detailed and quantitative comparison study of the calculated spin
noise was performed using experiments in a classical gas of
$^{41}$K atoms, and we report good agreement between theory and
experiment in both longitudinal and transverse applied magnetic
fields, from low fields up to the high-field regime where Zeeman
energies are comparable with hyperfine energies. The theoretical
results presented here apply to both classical gases at high
temperature as well as ultracold quantum gases. Because the
integrated strength of the lines gives information about the
occupation of the atomic levels (while the line-shapes depend on
the properties of the condensed atomic state) future spin noise
spectroscopy measurements may play an important role for the study
of the effective interaction in ultracold atom gases.


\begin{acknowledgments}
This work was supported in part by the LDRD program at Los Alamos
National Laboratory. B.M. acknowledges financial support from an
ICAM fellowship program. The authors would like to thank
A.V.~Balatsky for useful discussions.
\end{acknowledgments}


\end{document}